bi

\documentclass[letterpaper,twocolumn,10pt]{article}
\pdfoutput=1

\usepackage{usenix2019_v3}

\usepackage{tikz}
\usepackage{amsmath}
\graphicspath{ {figures/} }
\usepackage{subcaption}
\usepackage[labelfont=bf]{caption}
\usepackage{fmtcount} 
\usepackage{multirow}
\usepackage{multicol}
\usepackage{booktabs}

\usepackage{array}
\newcolumntype{C}[1]{>{\centering\arraybackslash}m{#1}}
\newcolumntype{L}{>{\centering\arraybackslash}m{3cm}}
\usepackage[title]{appendix}
\usepackage{color}

\renewcommand{\paragraph}[1]{\vspace{0.04in}\noindent{\bf{#1}.}} 
\begin{document}

\date{}

\title{\Large \bf An Object Detection based Solver for Google's Image reCAPTCHA v2} 


\author{
{Md Imran Hossen$^*$ $\;\;$

Yazhou Tu$^*$ $\;\;$ 

Md Fazle Rabby$^*$ $\;\;$

Md Nazmul Islam$^*$ $\;\;$

Hui Cao$^\dag$$\;\;$   Xiali Hei$^*$} \\
$^*$University of Louisiana at Lafayette\\
$^\dag$Xi'an Jiaotong University
}




\maketitle

\begin{abstract}
Previous work showed that reCAPTCHA v2's image challenges could be solved by automated programs armed with Deep Neural Network (DNN) image classifiers and vision APIs provided by off-the-shelf image recognition services. 
In response to emerging threats, Google has made significant updates to its image reCAPTCHA v2 challenges that can render the prior approaches ineffective to a great extent.  
In this paper, we investigate the robustness of the latest version 
of reCAPTCHA v2 against advanced object detection based solvers. We propose a fully automated object detection based system that breaks the most advanced challenges of reCAPTCHA v2 with an online success rate of 83.25\%, the highest success rate to date, and it takes only 19.93 seconds (including network delays) on average to crack a challenge. We also study the updated security features of reCAPTCHA v2, such as anti-recognition mechanisms, improved anti-bot detection techniques, and adjustable security preferences. 
Our extensive experiments show that while these security features can provide some resistance against automated attacks, adversaries can still bypass most of them. 
Our experimental findings indicate that the recent advances in object detection technologies pose a severe threat to the security of image captcha designs relying on simple object detection as their underlying AI problem.
\vspace{-1mm}
\end{abstract}
\vspace{-1mm}

\section{Introduction} \vspace{-1mm}

CAPTCHA is a defense mechanism against malicious bot programs on the Internet by presenting users a test that most humans can pass, but current computer programs cannot \cite{vonAhn_2004}.  
Often, CAPTCHA makes use of a hard and unsolved AI problem.
Over the last two decades, text CAPTCHAs have become increasingly vulnerable to automated attacks as the underlying AI problems have become solvable by computer programs \cite{Mori:2003:ROA:1965841.1965858, Moy:2004:DET:1896300.1896305, Chellapilla:2004:UML:2976040.2976074, Yan2007BreakingVC, Yan2007BreakingVC, Yan_2008, Bursztein_CCS11, Gao_13CCS, Bursztein_2014, Gao2016ASG, Ye_2018}. As a result, text CAPTCHAs are no longer considered secure. 
In fact, in March 2018, Google shut down its popular text CAPTCHA scheme reCAPTCHA v1 \cite{recaptchav1shutdown}. Image CAPTCHA schemes have emerged as a superior alternative to text ones as they are considered more robust to automated attacks. 

reCAPTCHA v2, a dominant image CAPTCHA service released by Google in 2014, asks users to perform an image recognition task to verify that they are humans and not bots. However, in recent years, deep learning (DL) algorithms have achieved impressive successes in several complex image recognition tasks, often matching or even outperforming the cognitive ability of humans \cite{He_2015}. Consequently, successful attacks against reCAPTCHA v2 that leverage Deep Neural Network (DNN) image classifier and off-the-shelf (OTS) image recognition services have been proposed \cite{Sivakorn_2016, Weng_2019}. 

The prior work advanced our understanding of the security issues of image CAPTCHAs and led to better CAPTCHA designs. 
However, recently, Google has made several major security updates to reCAPTCHA v2 image challenges that can render prior image classification and recognition based approaches ineffective to a great extent. For example, the latest version of reCAPTCHA pulls challenge images from relatively complex and common scenes as opposed to monotonic and simple images in the past. 
Through a comprehensive experiment, we show that both image classifiers and image recognition APIs provide poor success rates against the latest reCAPTCHA v2 challenges. 


Our experiment also shows that the current version of reCAPTCHA v2 adopts several additional security enhancements over the earlier versions. First, reCAPTCHA v2 has introduced anti-recognition techniques to render the challenge images unrecognizable to state-of-the-art image recognition technologies. For example, it often presents noisy, blurry, and distorted images. reCAPTCHA image challenges are likely to be using adversarial examples \cite{szegedy2013intriguing, Biggio_2013} as a part of the anti-recognition mechanism as well. Second, it adapts the difficulty-level for suspicious clients by presenting them with harder challenges. Third, the improved anti-bot detection mechanism of reCAPTCHA can now detect the popular web automation framework like Selenium.
Apart from those, reCAPTCHA v2 also added click-based CAPTCHA tests, which are not explored in the prior studies. We suspect that the click-based CAPTCHAs were not available at the time of publication of the most recent attack on reCAPTCHA v2. 

Taking reCAPTCHA v2 as an example, we investigate the security of image CAPTCHA schemes against advanced object detection technologies. To this end, we develop an object detection based real-time solver that can identify and localize target objects in reCAPTCHA's most complex images with high accuracy and efficiency. Specifically, our system can break reCAPTCHA image challenges with a success rate of 83.25\%, the highest success rate to date, and it takes only 19.93 seconds (including network delays) on average to crack a challenge. 
Our economic analysis of human-based CAPTCHA solving services shows that our automated CAPTCHA solver provides comparable performance to human labor. Therefore, the scammers can exploit our system as an alternative to human labor to launch a large-scale attack against reCAPTCHA v2 for monetary or malicious purposes, leaving millions of websites at the risk of being abused by bots \cite{recaptcha_v2_UsageStats}.

We also provide an extensive analysis of the security features of the latest version of reCAPTCHA v2. First, we find that the anti-recognition mechanisms employed by reCAPTCHA can significantly degrade the performance of both image recognition and object detection based solvers. However, our extensive analysis shows that we can neutralize reCAPTCHA's anti-recognition attempts by applying advanced training methods to develop a highly effective object detection based solver. Second, we also find that our system can bypass many other imposed security restrictions. For example, we can bypass the browser automation framework restriction by using the puppeteer-firefox  \cite{puppeteer_for_firefox} framework. 
Our findings reveal that despite all the evident initiatives
by Google, reCAPTCHA still fails to meet the stringent security requirements of a secure and robust CAPTCHA scheme. 







In summary, we make the following contributions:
\vspace{-3mm}

\begin{itemize}

\item Through extensive analysis, we show that prior DNN image classifiers and off-the-shelf vision APIs based approaches are no longer effective against the latest version of reCAPTCHA v2.
We then propose an object detection based attack that can break the most advanced image challenges provided by reCAPTCHA v2 with high accuracy and efficiency. 
%
%

\vspace{-2mm}


\item We provide a comprehensive security analysis of different security features employed by the latest version of reCAPTCHA v2. Our extensive study shows that these features can provide some resistance to automated attacks. However, adversaries can still bypass most of them. 
\vspace{-2mm} 



\item
Our study indicates that the recent advances in object detection algorithms can severely undermine the security of image CAPTCHA designs. As such, the broader impact of our attack is that any image CAPTCHA schemes relying on simple object detection as their underlying AI problem to make a distinction between bots and humans might be susceptible to this kind of attack. 

\end{itemize}

 \begin{figure}[htp]
\centering
 \includegraphics[scale=0.17]{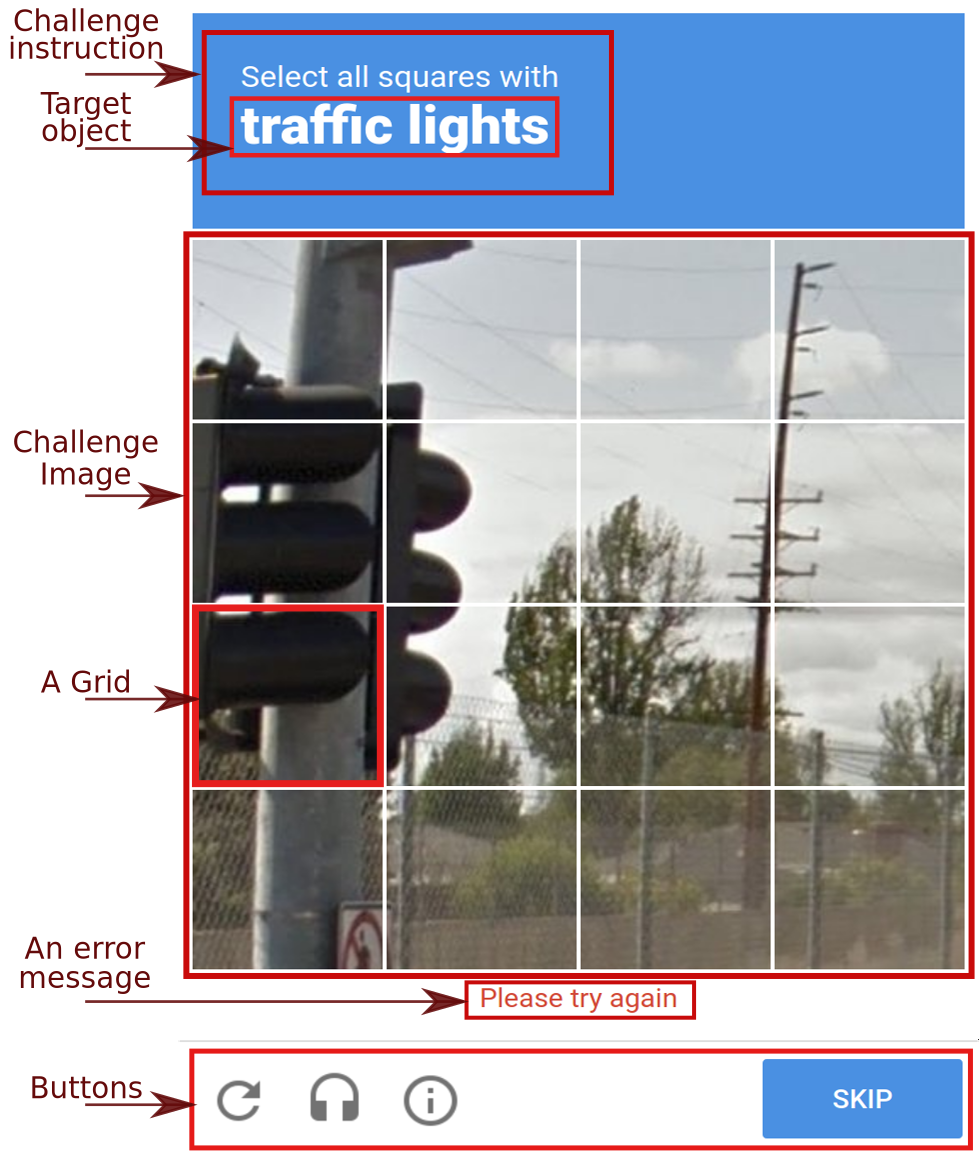}
 \caption{A reCAPTCHA v2 challenge widget.}\label{fig:challenge_widget}
\end{figure}

\vspace{-2mm} 
\section{reCAPTCHA v2 background} \label{sec:background} \vspace{-2mm}

reCAPTCHA v2 relies on an \textit{advanced risk analysis engine} to score users' requests and let legitimate users bypass the CAPTCHA test. Once the user clicks the reCAPTCHA -- ``I'm not a robot'' --- checkbox, the advanced risk analysis engine tries to determine whether the user is a human using various signals collected by the system, including different aspects of the user's browser environment, and Google tracking cookies \cite{InsideReCaptcha, Sivakorn_2016}. If the system finds the user suspicious, it asks the user to solve one or more image CAPTCHA(s) to prove that he/she is a human and not a bot. In general, a user with no history with Google services will be assigned to relatively difficult challenges. In this paper, our system attempts to solve these CAPTCHAs. Note that, Bock \textit{et al}. followed a similar approach to break reCAPTCHA's audio challenges in 2017 \cite{Bock:2017:ULD:3154768.3154775}. 


It is important to note that the third version of reCAPTCHA, reCAPTCHA v3, was released in October 2018. reCAPTCHA v3 is intended to be frictionless, \textit{i.e.}, not requiring any users' involvement in passing a challenge. However, it has raised some serious security concerns due to the method it uses to collect users' information \cite{rcV3_Issue_Regsiter, rcV3_darkside}. In this paper, we only target reCAPTCHA v2's most recent (as of March 2020) image challenges because it is still the most popular and widely used version of reCAPTCHA deployed on the Internet. From now on, we will use the term reCAPTCHA to refer to reCAPTCHA v2 unless otherwise specified. 


\paragraph{Challenge widget} If reCAPTCHA requires the user to solve a challenge, a new \texttt{iframe} gets loaded on the webpage after clicking on the ``I'm not a robot'' checkbox. The \texttt{iframe} contains the actual reCAPTCHA challenge (Figure \ref{fig:challenge_widget}). The challenge widget can be divided into three sections: top, middle, and bottom. The top section includes instructions about how to solve the challenge.
The section in the middle holds the candidate images. The user has to select or click on images that contain the target object mentioned in the instruction. At the bottom, it has multiple buttons, including the ``reload'' button, ``audio CAPTCHA'' button, and the ``verify'' button. 
\begin{figure}[!t]  
    \centering
    \includegraphics[scale=0.20]{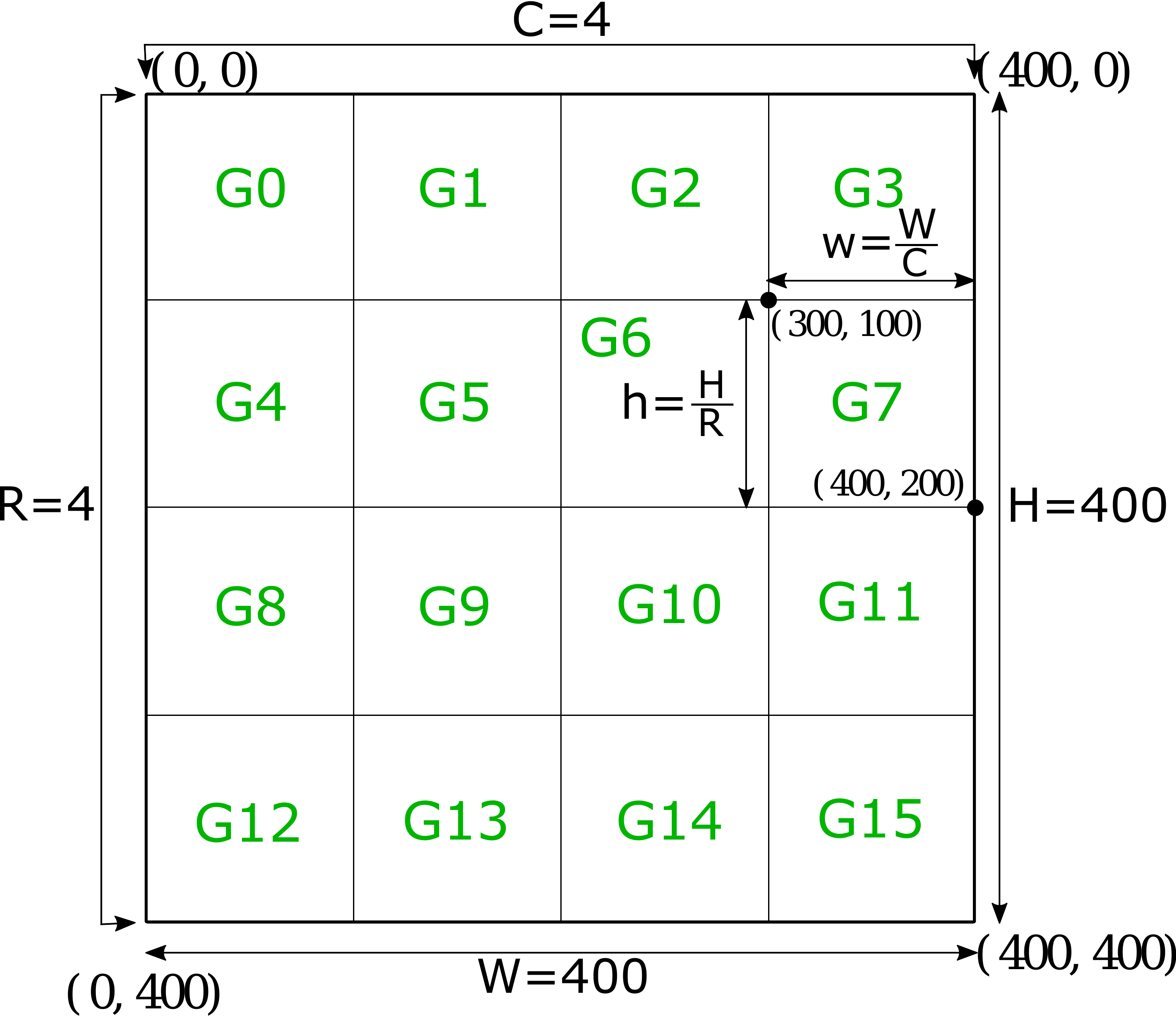}
    \caption{Representing a 400px $\times$ 400px challenge image as an $R \times C$ grid. Here, $R$=No. of rows, and $C$=No. of cells per row in the HTML table element holding the challenge image.}
    \label{fig:grid_repres}
\end{figure}



The images are located inside an HTML table element. The table has multiple rows, and each row holds the same number of cells. Each cell has an \texttt{img} tag in it and renders an image from the URL specified in the tag. If the table has 4 rows and each row has 4 cells, then 16 candidate images in total will be rendered, from which the user has to select the right images. However, all these images are pulled from a single source URL in the challenge widget initially, \textit{i.e.}, a single image is split across multiple table cells in an equal proportion. Therefore, this particular challenge image can be treated as a 4 $\times$ 4 grid. Figure \ref{fig:grid_repres} illustrates the process of representing a 400px $\times$ 400px challenge image as an $R \times C$ grid. Here $R$ and $C$ correspond to the number of rows and the number of cells per row in the table element. 

\paragraph{reCAPTCHA CAPTCHA types} The current version of reCAPTCHA has two types of image CAPTCHAs: 1) \textit{selection-based image CAPTCHA} and 2) \textit{ click-based image CAPTCHA}. The selection-based CAPTCHA requires the user to select the correct grids containing the right object as specified in the instruction 
to pass the challenge (see Figure \ref{fig:sel_captcha} in Appendix \ref{appendix:captchaTypes}). It is the common CAPTCHA type that the user would encounter in a reCAPTCHA-protected site. The click-based CAPTCHAs have been introduced only recently. In a click-based image challenge, when the user clicks on a grid, the image on the grid disappears, and a new image gets generated in its place (see Figure \ref{fig:click_captcha} in Appendix \ref{appendix:captchaTypes}). The user has to repetitively click on the potential grids until the target object is no longer present in any of the grid while submitting a challenge. 
It takes a relatively long time to solve the click-based CAPTCHAs than selection-based ones as there is a delay between the click and image regeneration process. 
\vspace{-2mm}
\section{Threat model}  \vspace{-2mm}
We assume the attacker's goal is to abuse Web applications protected by reCAPTCHA using an automated program. We also assume the attacker has access to a GPU enabled machine to deploy an object detection system for cracking CAPTCHAs. The attacker can launch the attack from a single IP address. However, having access to a large IP pool will allow the attacker to launch a large-scale attack. reCAPTCHA may occasionally block an IP address for some time. In such a scenario, the attacker may need to use a proxy service or anonymity network such as Tor \cite{Dingledine:2004:TSO:1251375.1251396} to bypass the IP restriction. In summary, we consider a low-to-moderately resourced attacker whose goal is to deploy a highly effective automated solver to break reCAPTCHA challenges for malicious purposes. 

\vspace{-2mm}
\section{Our approach} \label{sec:overview} \vspace{-2mm}

Our automated CAPTCHA breaker consists of a browser automation module and a solver module. 

\paragraph{Browser automation module} The \emph{browser automation module} is responsible for automating different browser-specific tasks while solving a CAPTCHA challenge. These tasks include locating the reCAPTCHA checkbox, initiating the reCAPTCHA challenge, and identifying the potential HTML elements on the challenge widget. This module is also in charge of fetching challenge images, submitting the solution once the CAPTCHA is solved, monitoring the progress of the challenge, and checking the reCAPTCHA verification status. 

\paragraph{Solver module} The solver module consists of two main components: the base object detector and the \textit{bounding box to grid mapping} algorithm. The base object detector takes the challenge image from the \emph{browser automation module} and identifies and localizes objects in it. For each recognized object instance, the base detector returns its class name, the confidence score, and coordinate information in terms of the bounding box. The \textit{bounding box to grid mapping} algorithm then uses this data to map the bounding boxes holding the target object back to the grids where they are present. 


\subsection{The browser automation module} 
The \emph{browser automation module} first visits a reCAPTCHA-protected webpage and locates the frame element holding ``I am not a robot'' checkbox. It then clicks on the checkbox, which is identified by \texttt{recaptcha-anchor}, to initiate the challenge. 
Now our system switches to the challenge widget. Then it primarily conducts the following steps to solve the challenge. 

\paragraph{Extracting challenge instruction} Our system locates the element holding the challenge instruction \texttt{rc-imageselect-instructions}. The challenge instruction is a multi-line string, and the second line always refers to the name of the target object. Further, it indicates the challenge type. For instance, in click-based image CAPTCHAs, the challenge instruction always holds the phrase --- ``Click verify once there are none left'' (See Figure \ref{fig:click_captcha}). The name of the target object, for which we must solve the challenge, may not always be in the singular form. If that happens, we singularize it.  

\paragraph{Determining the total number of rows ($R$) and the number of cells per rows ($C$)} As discussed in Section \ref{sec:background}, if we know the total number of rows ($R$) and the number of cells per row ($C$) in the HTML table holding the challenge image, we can represent the challenge image as an $ R \times C $ grid. We use JavaScript methods to determine $R$ and $C$. 

\paragraph{Downloading challenge image}
The element \texttt{rc-imageselect-tile} holds the challenge image. There will be multiple such elements based on the total number of grids. Since all the elements link to the same image, our system downloads the first image only. However, for click-based CAPTCHAs, it will need to download dynamically loaded images on the selected grids as well.

\paragraph{Identifying buttons on the challenge widget} To submit the challenge once it is solved, we need to click on the ``verify'' button. Our system locates the ``verify'' button using its identifiers \texttt{recaptcha-verify-button}.

\subsection{Implementation of the solver module}  
The solver module identifies and localizes target objects in reCAPTCHA challenge images. Further, the module is responsible for mapping the detected objects back to their corresponding (potential) grids in the original challenge. The two main components of this module are a \textit{base object detection system} and the \textit{bounding box to the grid mapping} algorithm. We now discuss each of them in detail. 

\subsubsection{Base object detector: YOLOv3} 

We use YOLOv3 as the base object detector after experimenting with several other advanced object detectors, including Faster R-CNN \cite{Ren_2017}, R-FCN \cite{dai2016rfcn}, SSD \cite{Liu_2016_SSD}, and RetinaNet \cite{Lin_2017}. We find YOLOv3 to be significantly faster than all other tested object detectors when running the detection on a test image; however, the accuracy of YOLOv3 might be slightly lower than other object detectors. Since solving CAPTCHAs is a time-sensitive task, we opt to use YOLOv3 for its superior speed. The feature extractor network in YOLOv3 is called Darknet-53 because it has 53 convolutional layers, with shortcut connections.
See \cite{yolov3} for details.

\paragraph{Datasets} We use two datasets, specifically developed to handle object categories found in reCAPTCHA challenges. The first dataset is a publicly available dataset called MS COCO \cite{mscoco}. The MS COCO dataset has 80,000 training images and 40,000 validation images with 80 object classes, out of which 8 classes frequently appear in reCAPTCHA challenges. The MS COCO object classes common to reCAPTCHA object categories are bicycle, boat, bus, car, fire hydrant, motorcycle, parking meter, and traffic light. The second dataset is a custom one that we develop by ourselves. We crawled over 6,000 images from different sources such as Flickr  \footnote{https://www.flickr.com/}, Google image search \footnote{https://images.google.com/}, and Bing image search \footnote{https://www.bing.com/images/}. After prepossessing these, we end up with 4800 images. We also use 2100 images from the original reCAPTCHA challenges for this dataset. We manually annotated and labeled the object instances in those images to prepare and finalize the dataset. Our final custom dataset has 11 object categories: boat, bridge, chimney, crosswalk, mountain, palm tree, stair, statue, taxi, tractor, and tree. 

\paragraph{Training the base object detector} We use two YOLOv3 models trained on the two datasets. We mostly go with the default architecture of the YOLOv3 network with some minor modifications for both models. We set up the batch size to 64, and the learning rate to 0.001 for training. We use the Darknet \cite{darknet13}, an open-source neural network framework written in C, for training the YOLOv3 models. We train the model on the MS COCO dataset for roughly 15 days, and the model on the custom dataset for 2 days. The training is performed on a server with an NVIDIA RTX 2070 GPU. We then evaluate the weight files for both models on corresponding test sets and choose the best weights. Our final model for the MS COCO dataset has the mean average precision at 0.5 IOU (mAP@.5) of 57.4\% on the testing set. The second model has obtained a mAP@.5 value of 51.79\% on the respective testing set. 

\paragraph{Inference or making predictions} We use the Darknet framework to make predictions on reCAPTCHA challenge images with our trained models. By default, Darknet does not provide any localization information. We adjust the source code to output the bounding box coordinates when running the inference on an image. The modified prediction output includes class name, confidence score, and bounding box coordinates for each detected object instance in a prediction operation. We set the detection threshold to 0.2. 

\subsubsection{The bounding box to grid mapping algorithm}  

After detecting the objects with the base object detector in the challenge image, we need to map the objects back to their corresponding grids in the original challenge. Our \textit{bounding box to grid mapping} algorithm works as follows.

\begin{enumerate}
  \item Use the $R$ and $C$ parameters from the \emph{browser automation module} to get an $ R \times C $ grid representation of the image.  
  \item Compute coordinates of each grid relative to the top left of the image (see Figure \ref{fig:grid_repres}).  
  \item Take the prediction output from the base object detector.   
  \item For each bounding box with the class label matching the target object name in the challenge, take the coordinates of the box and the grids. If any of the coordinates of the bounding box falls inside a grid, mark it as a potential grid. Depending on the size of the bounding box, it may fall within multiple grids. We store all of these grid numbers in an array and call it as the potential grid numbers (PGNs).  
  \item For each bounding box, return the class name, confidence score, and the PGNs.  
  \item Return the results as a JSON array.  
\end{enumerate} 

\begin{figure}[!t]
    \includegraphics[width=1.0\linewidth, height=5cm]{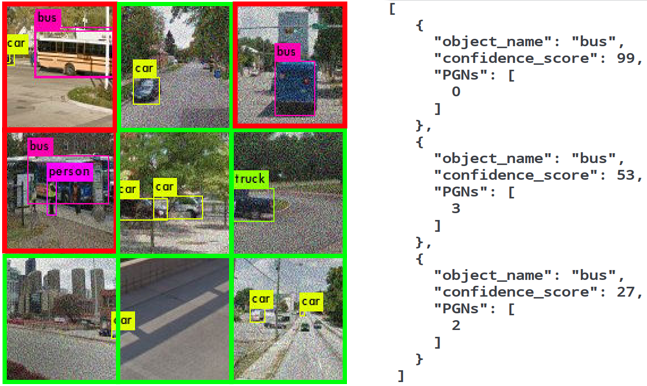}
    \caption{A result returned by the \textit{bounding box to grid mapping} algorithm. Left: An original challenge image (the target object is a ``bus''). Right: The JSON array returned by the algorithm.}
    \label{fig:bbox_to_grid_mapping} 
\end{figure}

Figure \ref{fig:bbox_to_grid_mapping} shows an example of the result returned by the \textit{bounding box to the grid mapping} algorithm for a sample challenge image. 

\vspace{-2mm}
\subsection{Submitting and verifying challenges} 
The JSON array returned by the solver module is passed to the \emph{browser automation module}. The browser automation module first extracts the potential grid numbers from the PGNs arrays and locates the representative grids in the HTML table. It then clicks on these grids in the challenge widget and finally clicks the ``verify'' button when the process is completed. 

The system then verifies whether the challenge is passed or not using the ``reCAPTCHA ARIA status messages.'' \footnote{https://support.google.com/recaptcha/\#aria\_status\_message} We further verify the challenges submitted to our own websites by validating user response token, \texttt{g-recaptcha-response}, to the reCAPTCHA backend. The \texttt{g-recaptcha-response} remains empty until the challenge is solved. When a challenge is successfully solved, it gets populated with a long string.  
After submitting a challenge to our website, our bot first extracts the user response token. It then sends a verification request to the reCAPTCHA backend server with this token and the secret key to authenticate the token.


\begin{figure}[!t]
    \includegraphics[width=0.9\linewidth, height=7cm]{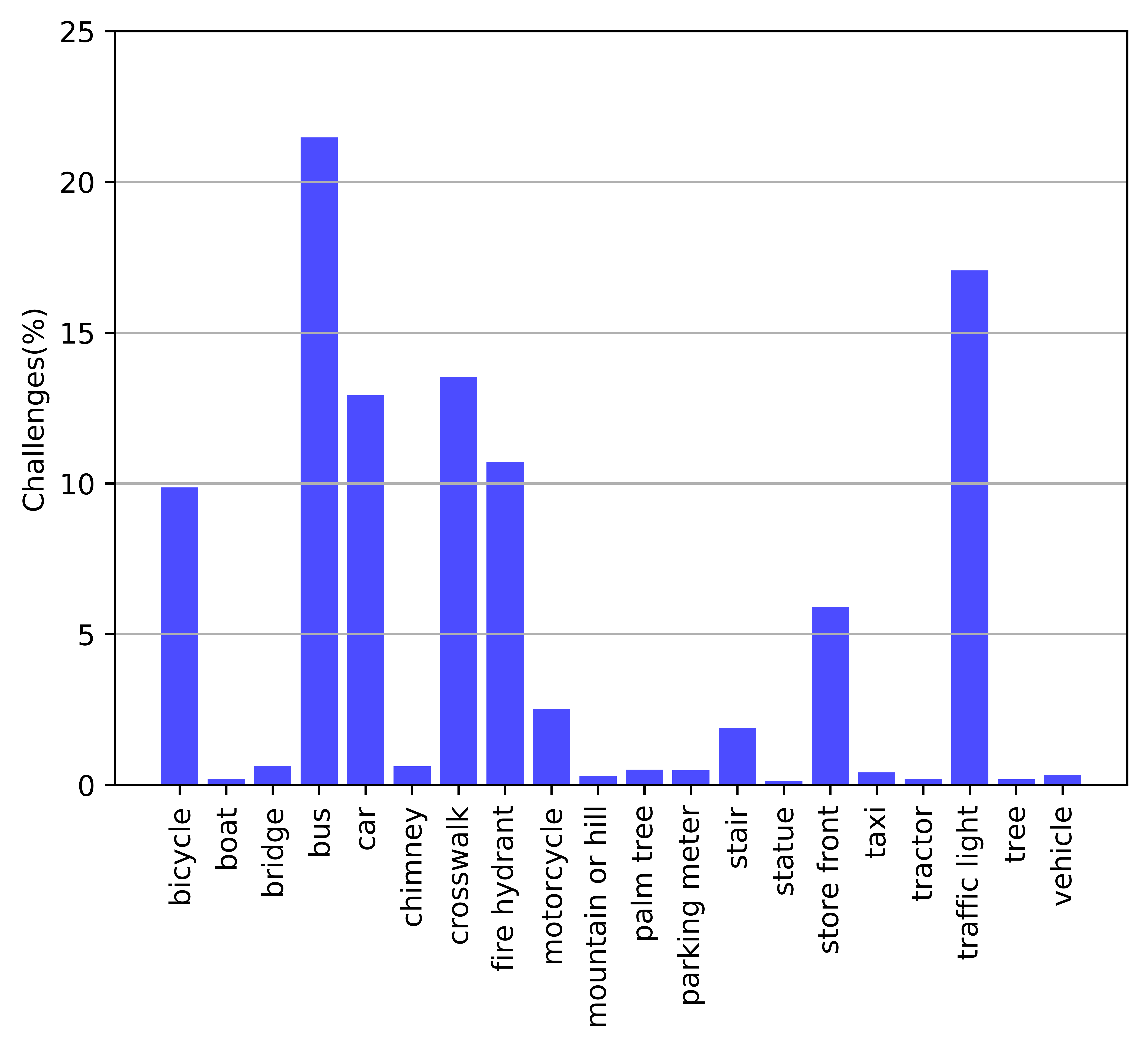}
    \caption{The frequency of different object categories in collected challenges.}
    \label{fig:prelim_object_vs_freq}
\end{figure}

\vspace{-2mm}
\section{Attack evaluation} \vspace{-2mm}

\subsection{Implementation and evaluation platform} \label{subsec:eval_platform} 

The \emph{browser automation module} is built upon the puppeteer-firefox \cite{puppeteer_for_firefox}, a node library developed by Google, to control the Firefox web browser programmatically. The core functionalities of the module are developed using JavaScript. The base object detector in the solver module is based on the YOLOv3 object detection algorithm. We train and test the YOLOv3 models with a customized version of the Darknet framework that especially meets our needs. Our \textit{bounding box to the grid mapping} algorithm is written in C for efficiency. 

We train the YOLOv3 models on a server with 6 Intel\textsuperscript{\tiny\textregistered} Xeon\textsuperscript{\tiny\textregistered} E5-2667 CPUs, an NVIDIA GeForce RTX 2070 GPU, and 96GB of RAM running the Arch Linux operating system. We compile the Darknet framework against the GPU with CUDA 10.2 and cuDNN 7.5. We conduct all the experiments on this machine. 

\paragraph{Experimental setting} We use puppeteer-firefox (version 0.5.1) running on top of node library version 14.4.0 for browser automation. The browser we used is Firefox 65.0. We run the web browser in a clean state each time, \textit{i.e}., no caches or cookies are retained between two subsequent requests. We do not attempt to obfuscate any aspects of the requests or web browser properties (e.g., using custom \texttt{User-Agent} header or modifying the \texttt{Navigator} object properties). 
Further, unless otherwise specified, all the requests to reCAPTCHA-protected websites are made from a single IP address and a single machine.
\vspace{-2mm}
\subsection{Preliminary analysis} \vspace{-2mm}

To do the preliminary analysis, we collect 10,385 reCAPTCHA challenges from 8 websites protected by reCAPTCHA service from May 2019 to November 2019. Figure \ref{fig:prelim_object_vs_freq} shows frequencies of different object categories in the collected challenges. As we can see, there are only 19 object categories, with the top 5 categories representing over 75\% of the total challenges. The top 5 object classes are bus, traffic light, crosswalk, car, and fire hydrant.

\begin{table}[!t] 
\footnotesize
    \caption{Performance of image recognition services.} 
    \label{tab:perf_vision_api}
    \begin{tabular}{ L|c|c }
    \toprule
     \textbf{Image Recognition Service} & \textbf{Success Rate (\%)} & \textbf{Speed (s)} \\
     \hline
     Google Cloud Vision & 19 & 16.67 \\
     \hline
     Microsoft Azure Computer Vision & 34 & 17.90 \\
     \hline
     Amazon Rekognition & 47 & 19.68 \\
     \hline
     Clarifai & 27 & 15.35 \\
     \bottomrule
    \end{tabular} 
\end{table}

\begin{table}[!t] 
\footnotesize
    \caption{Performance of object detection services.} 
    \label{tab:perf_vision_api_obj_det}
    \begin{tabular}{ L|c|c }
    \toprule
     \textbf{Object Detection Service} & \textbf{Success Rate (\%)} & \textbf{Speed (s)} \\
     \hline
     Google Cloud Vision & 36 & 9.47 \\
     \hline
     Microsoft Azure Computer Vision & 31 & 10.54 \\
     \hline
     Amazon Rekognition & 38 & 9.86 \\
     \bottomrule
    \end{tabular} \vspace{-3mm}
\end{table}

\paragraph{Breaking reCAPTCHA with vision APIs for image recognition}
We test 4 popular off-the-shelf online vision APIs for image recognition. The services we use are Cloud Vision API provided by Google \cite{vision_api_google}, Azure Computer Vision API provided by Microsoft \cite{vision_api_microsoft}, Rekognition API provided by Amazon \cite{vision_api_amazon}, and the API provided by Clarifai \cite{vision_api_clarifai}.
First, we select some challenge images from different categories, extract the individual grids from them, and submit those grids to the image recognition services to analyze the tags (labels) returned by them. We find that in most cases, one of the labels for a grid holding the target object matches precisely with the name of the object in the reCAPTCHA challenge instruction, thus simplifying the process of mapping the tags returned by an API service to reCAPTCHA challenge object names while submitting a challenge. 
However, we find one instance where the labels are not consistent across various APIs. For example, Amazon Rekognition API classifies reCAPTCHA's crosswalk images as ``Zebra Crossings,'' while Google's Cloud Vision API recognizes them as ``Pedestrian crossings.'' We do a simple preprocessing that transforms these labels to name ``crosswalk'' for consistency.

Next, we develop a proof-of-concept attack and submit 100 live reCAPTCHA challenges separately using each image recognition API. Table \ref{tab:perf_vision_api} provides the success rate and speed of attack using image recognition services. The Google Cloud Vision provides the lowest success rate, followed by Clarifai. We note that the attack success rate is below 40\% for all the services except for Amazon Rekognition. 

Finally, we manually verify the results and analyze the failed challenges. We find that the image recognition services' poor performance is due to the complex nature of the current challenge images, which often contain complex everyday scenes with common objects in their natural context. For example, we find many instances where a potential grid holding a ``crosswalk'' also holds other common objects such as ``car'' and ``traffic light'' in it, and tags returned by an API include names of all the objects except the primary target.
Further, in many challenges, a single target object spans across multiple grids, and some of those grids contain only a tiny part of the whole object. Image recognition services failed to identify the target object in such a scenario. The earlier version of reCAPTCHA used to show relatively simple images, usually containing one disparate object per grid or images with simple scenes having a monotonic background, making it easier for image recognition services to analyze the contents. 

\paragraph{Breaking reCAPTCHA with an image classifier}
We also perform an attack using a Convolutional Neural Network (CNN) based image classifier. The classifier is trained on over 98,000 images from 18 classes. These include all object classes in Figure \ref{fig:prelim_object_vs_freq}, except the ``store front.'' 
Interestingly, the ``store front'' class has been phased-out from reCAPTCHA challenges during the later part of our data collection period. We then submit 100 live reCAPTCHA challenges using our image classifier based solver. The success rate and speed of attack are 21\% and 16.96 seconds, respectively. After analyzing the failed challenges manually, we find that the same factors related to the poor performance of the image recognition APIs contributed equally (or even higher) to the low success rate of the image classifier based attack. 

\paragraph{Breaking reCAPTCHA with online vision APIs for object detection}
We also carry out a proof-of-concept attack using three off-the-shelf computer vision APIs for object detection provided by Google, Microsoft, and Amazon. We customize our \textit{bounding box to the grid mapping} algorithm to process the bounding box results for the objects detected by the APIs. Like before, we submit 100 live reCAPTCHA challenges using these APIs. Table \ref{tab:perf_vision_api_obj_det} shows attack performance of each off-the-shelf object detection API. We can see that the Amazon Rekognition API and Google Cloud Vision API achieve similar performance, while Microsoft Azure Computer Vision API performs relatively poorly. We analyze the results to understand why these services are not effective against reCAPTCHA challenges. We find several factors that contribute to low success rates. First, these services can recognize objects from certain object categories only. However, most of them can detect objects that frequently appear in the reCAPTCHA challenges such as ``bus'', ``car'', ``traffic light'', and ``bicycle''. While the top 5 objects in Figure \ref{fig:prelim_object_vs_freq} account for around 70\% of the submitted challenges during this experiment, our manual analysis shows that the object detection APIs fail to identify at least one target object in these categories in most of the failed cases. It suggests that the cloud-based vision APIs for object detection are still in their early stage of development, and yet to be ready to handle complex images such as those found in reCAPTCHA challenges. 

\subsection{Breaking reCAPTCHA challenges with our system} 

\paragraph{Success rate and speed of attack}
To evaluate the effectiveness and efficiency of our approach, we submit 800 challenges to 4 reCAPTCHA-enabled websites using our automated CAPTCHA-breaking system. Out of them, 701 challenges are selection-based, 87 challenges are click-based, and 12 are ``no CAPTCHA reCAPTCHA'' challenges, where our system gets verified simply by clicking on the reCAPTCHA checkbox. Our system breaks 656 (out of 788) challenges, resulting in a success rate of 83.25\%.

The average speed of breaking a CAPTCHA challenge is 19.93 seconds, including delays. The delays include network delay to load and download images, which takes about 1\char`\~8 seconds depending on CAPTCHA types and artificially induced delay between each of the clicks.
The minimum, median, and maximum time needed to break a challenge are 13.11, 14.92, and 89.02 seconds respectively. Table \ref{tab:percentiles} lists the time required to break a challenge by percentiles. 
Our solver module takes about 6.5 seconds to detect objects in a challenge image regardless of the number of objects being present. 

\begin{table}[!tp]
\centering
\caption{The time required to break a reCAPTCHA challenge by percentiles.}
\label{tab:percentiles}
\begin{tabular}{c|c|c|c|c}
\toprule
  \textbf{Percentile} & 1st & 5th  & 95th  & 99th  \\
  \hline
  \textbf{Speed (s)} & 13.28 & 13.91  & 39.76  & 58.65 \\
\bottomrule  
\end{tabular}
\end{table} 

\begin{figure}[!t]
\begin{subfigure}{.5\textwidth}
  \centering
  \includegraphics[width=.98\linewidth, height=6.5cm]{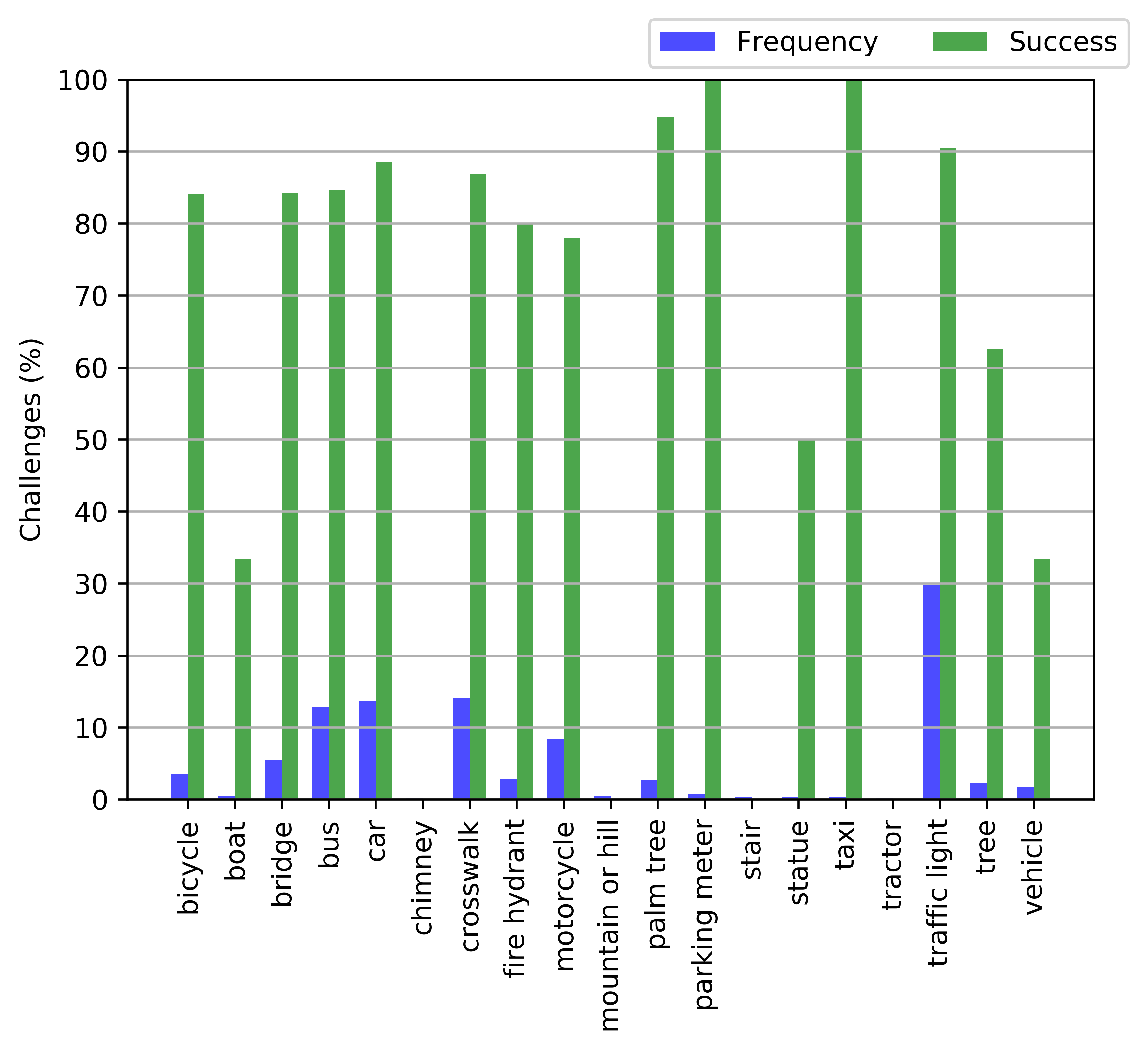}
  \caption{Selection-based CAPTCHA.}
  \label{fig:selection_attack_perf}
\end{subfigure}
\begin{subfigure}{.5\textwidth}
  \centering
  \includegraphics[width=.98\linewidth, height=5.5cm]{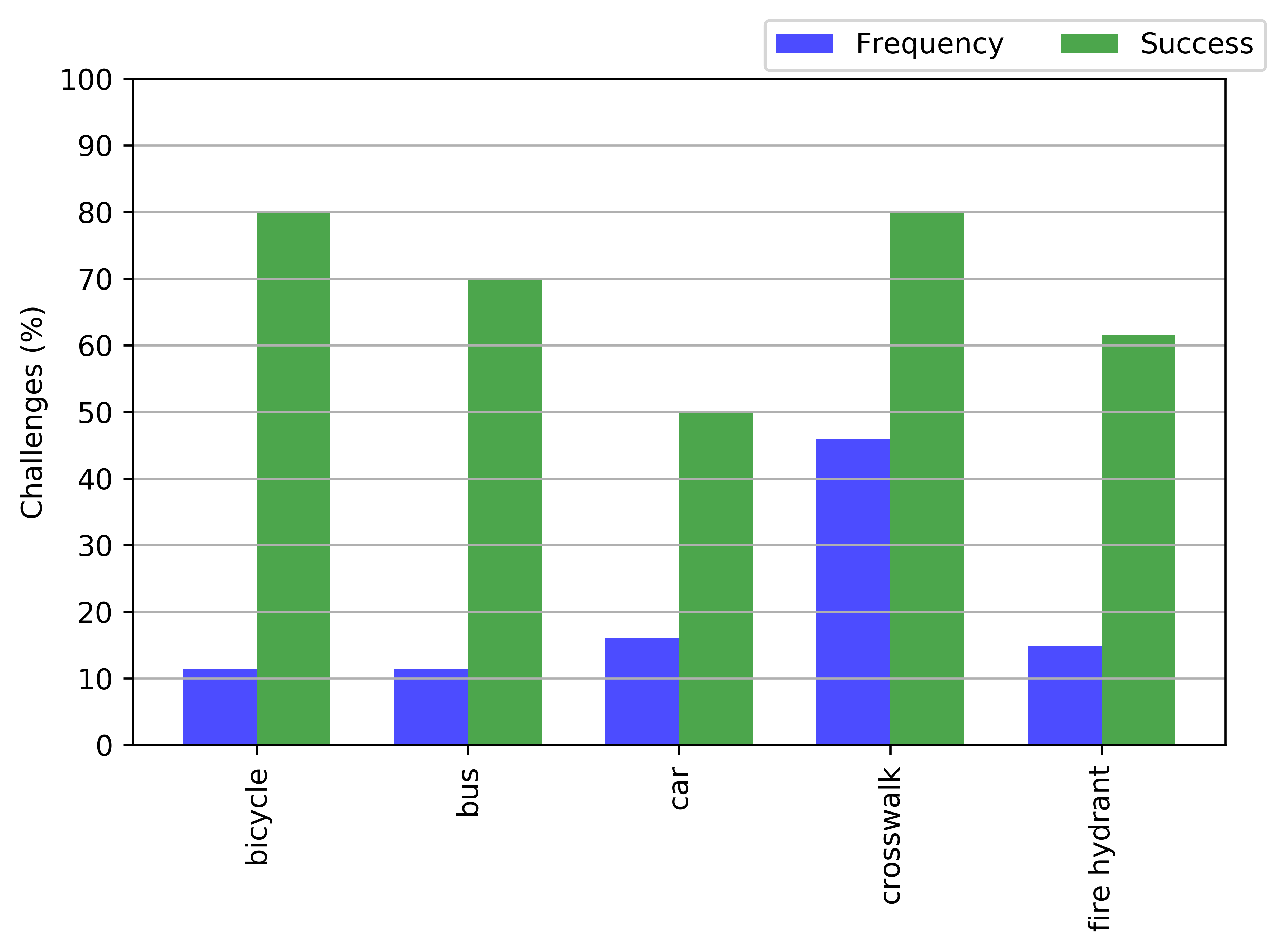}
  \caption{Click-based CAPTCHA.} 
  \label{fig:click_attack_perf}
\end{subfigure} 
\caption{Attack performance. Frequency and success rate for each object category.} 
\label{fig:attack_perf}
\end{figure}

\paragraph{Attack on selection-based CAPTCHAs} 
Generally, the selection-based image reCAPTCHA challenges appear more often than click-based ones. The success rate and speed of breaking selection-based challenges are 84.74\% and 17.47 seconds, respectively. Figure \ref{fig:selection_attack_perf} shows frequency and success rate for each object category in the submitted challenges. As we can see, only 19 object categories have been repeated across all 701 selection-based reCAPTCHA image challenges. Further, the top 5 object categories constitute over 78\% of the total challenges. If we consider the first 10 categories, this number goes above 95\%. 

Note that reCAPTCHA often asks users to solve multiple image puzzles in a row to pass a single selection-based reCAPTCHA test. However, in 80.81\% of passed CAPTCHAs, our system is required to solve only one image test. In 16.84\% of challenges, it is required to solve 2 image puzzles. The maximum number of puzzles required to pass a test is 5, and it occurs only twice. 

We find that in the majority of cases, there are at least 3 potential grids required to be chosen to pass a selection-based CAPTCHA test. Precisely, in 5.72\% of passed CAPTCHAs, our system is asked to select 2 grids. In 42.59\% of solved CAPTCHAs, it is required to choose 3 grids. In 32.15\% of solved challenges, the system is required to select 4 grids. The number of selected potential grids in the remaining challenges ranges from 5 to 14. We also find  2 tests where our system had to choose 18 grids to pass the challenges. It takes 4.01 seconds to select a grid while solving a challenge, on average. 

\paragraph{Attack on click-based CAPTCHAs} 
We come across only 87 click-based CAPTCHAs in the 800 submitted challenges, and our system passes 62 of them. The success rate and speed are 71.26\% and 43.53 seconds, respectively. Figure \ref{fig:click_attack_perf} provides the frequency and success rate for different object categories in click-based reCAPTCHA challenges. As we can see, in click-based CAPTCHA challenges, there are only five object categories. 


Since reCAPTCHA's click-based CAPTCHAs are relatively new and have not been explored in the previous study, we experiment to analyze the security of it further. The study by Sivakorn \textit{et al}. revealed that the initial implementation of reCAPTCHA v2 used to provide flexibility in its selection-based challenges, \textit{i.e.}, it used to accept 1 or 2 wrong grid selection(s) along with a certain number of correct grid selections while solving a CAPTCHA test \cite{Sivakorn_2016}. We investigate whether reCAPTCHA provides such flexibility for click-based challenges. We submit some CAPTCHAs by clicking on a different combination of correct and wrong grids. We submit 50 challenges for each combination. The result of our experiment is summarized in Table \ref{tab:click_based:grid_selection_test}. The highest success rate we achieve is 6\% when we click 6 correct grids and 1 wrong grid while submitting the challenges. It suggests that reCAPTCHA does not usually provide any solution flexibility for click-based CAPTCHAs. 

\begin{table}[!htp]
\small
    \caption{The success rates of different combinations of correct and wrong grid selection in a click-based CAPTCHA solution.} 
    \small
    \label{tab:click_based:grid_selection_test}
    \begin{tabular}{ c|c|c }
    \toprule
     \textbf{\# of Correct Grid} & \textbf{\# of Wrong Grid} & \textbf{Success Rate} \\
     \hline
     \hline
     3 & 1 & 0.00\% \\
     \hline
     4 & 1 & 2.00\% \\
     \hline
     5 & 1 & 0.00\% \\
     \hline
     6 & 1 & 6.00\% \\
     \hline
     6 & 2 & 0.00\% \\
     \hline
     3 (out of 4) & 0 & 0.00\% \\
     \hline
     4 (out of 5) & 0 & 0.00\% \\
     \hline
     5 (out of 6) & 0 & 2.00\% \\
     \bottomrule
    \end{tabular} 
\end{table}

\paragraph{Impact of security preference}
reCAPTCHA allows the website owners to adjust the security level based on their needs while deploying it to their websites. There are three levels in the security preference setting from ``Easiest for user'' to ``Most secure.'' By default, reCAPTCHA uses the security level in the middle, which we call ``Medium secure.'' To investigate the impact of these settings on the attackers' success rate, we deploy reCAPTCHA on our website and submit 50 challenges for each security setting. We run the experiment from a network that is isolated from the network hosting our webserver to avoid any biases.
We further follow the same experimental setting mentioned in  \ref{subsec:eval_platform}. 
%
%

The results of our findings are summarized in Table \ref{tab:result:security_pref}. The difference in the accuracy of our system across three security preferences is negligible. We have not noticed any obvious pattern that can distinguish one security preference from others. However, for the ``Easiest for user'' setting, we find that reCAPTCHA occasionally accepts a solution even when our bot misses one potential grid containing the target object or clicks on a wrong grid along with the correct grid selections. Further, reCAPTCHA often requires the bot to solve multiple image puzzles in a single selection-based challenge when using the ``Most secure'' security preference on the reCAPTCHA admin panel for our website.

\begin{table}[!t]
\small
    \caption{The success rates of different security preferences in the reCAPTCHA deployment setting.} 
    \label{tab:result:security_pref}
    \begin{tabular}{ c|c|c }
    \toprule
     \textbf{Security Preference} & \textbf{Success Rate (\%)} & \textbf{Speed (s)} \\
    \hline
     Easiest for user  & 82  &  16.75 \\
     Medium secure     & 78  &  14.31 \\
     Most secure       & 84  &  18.79 \\
     \bottomrule
    \end{tabular} 
\end{table}

\paragraph{Impact of browser automation software}
To study the impact of different browser automation frameworks on the performance of the bots, we develop a bot using the Selenium \cite{Selenium}. Selenium is one of the most widely used browser automation frameworks, which was also used by prior arts. 
Selenium provides WebDriver for both Mozilla Firefox and Google Chrome web browsers.
In particular, we use Selenium Python bindings (version 3.141.0) with Python version 2.7.18 in this experiment. For web browsers, we select Firefox 65.0 and Chrome 78.0.3882.0. To keep the experiment consistent with our main attack, we run the program from the same machine, and we access reCAPTCHA-enabled websites from the same IP address. Further, we clear the caches and cookies each time we launch the program. 

First, we use Firefox WebDriver. We submit 100 CAPTCHAs and notice that most of our attempts fail to break them. Accurately, the system can solve only 32\% of the total submitted challenges. A careful examination of our system log reveals that reCAPTCHA rejects many of the potentially correct solutions. Further, 12 out of 100 requests have been blocked with the message --- ``We're sorry, but your computer or network may be sending automated queries. To protect our users, we can't process your request right now.'' Note that at the same time, we also run our original puppeteer-firefox based system and verify that it can normally solve the challenges. Next, we repeat the same experiment with Chrome WebDriver. We recognize a similar pattern as before: the success rate of breaking the CAPTCHAs is below 40\%. We also find that reCAPTCHA shows a significantly higher percentage of click-based CAPTCHAs when we use the Selenium. Specifically, more than 25\% of the challenges that our system attempt to solve are click-based ones. Furthermore, we also encounter many noisy images. 

Since reCAPTCHA's advanced risk analysis engine treats our Selenium based system as highly suspicious, we try to obfuscate the presence of Selenium and investigate whether the obfuscation could improve our attack performance. When using an automation framework, the browser is supposed to set \texttt{navigator.webdriver} property to ``true'' according to W3C specification. However, an adversary may not follow this specification in an attempt to hide the presence of WebDriver to dodge detection. To experiment with an attacker's perspective, we set this property to ``false.''
Moreover, we obfuscate different aspects of the browser environment, such as the user-agent string, the number of plugins used, the number of fonts, and screen resolution. We then run a series of experiments with these settings. However, none of our attempts have resulted in any noticeable difference.

\begin{table*}[!t]
\footnotesize
\caption{A noisy image from reCAPTCHA challenge and the labels returned by 4 image recognition services. The target object is a ``Fire Hydrant.'' PGNs=Potential Grid Numbers (1, 3, 4). } 
\label{tab:noisy_image_}
\begin{center}
\begin{tabular}{c|c|C{2.5cm}|C{2.5cm}|C{2.5cm}|C{2.5cm}}
\hline
\textbf{Image}  & \textbf{PGNs} & \textbf{Google Cloud Vision} & \textbf{Microsoft Azure Computer Vision} & \textbf{Amazon Rekognition} & \textbf{Clarifai} \\
\toprule
\multirow{3}{*}{\includegraphics[scale=0.14]{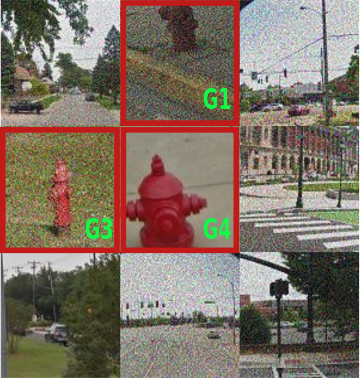}} &  1 &  leaf, grass, plant & animal, mammal  & art, painting  & desktop, animal, dog, people, nature  \\
 \cline{2-6}
 &  3 &  pattern, art & grass, outdoor  & hydrant, fire hydrant  & nature, abstract, pattern, art  \\
 \cline{2-6}
 &  4 &  pink, toy, action figure & hydrant, outdoor object, fire hydrant  & hydrant, fire hydrant & travel, old, art, desktop  \\
\bottomrule
\end{tabular} 
\end{center}
\end{table*}

\paragraph{Impact of anti-recognition}
We conduct a comprehensive analysis of the anti-recognition techniques employed by reCAPTCHA that have been introduced only recently in an attempt to undermine AI-based recognition. First, we need to identify noisy, distorted, or perturbed images. However, identifying such images is not trivial. A simple approach could be using Signal to Noise Ratio (SNR) measure to estimate the noise in an image. However, we cannot simply use SNR because we do not have access to the original ground truth images. In this situation, we find scikit-image's \cite{scikit-image} \texttt{estimate\_sigma} function to be particularly useful. \texttt{estimate\_sigma} provides a rough estimation of the average noise standard deviation ($\sigma_{noise}$) across color channels for an RGB image. We calculate $\sigma_{noise}$ for all 1,048 images that our solver attempt to solve during our experiment. The $\sigma_{noise}$ for 928 (over 88\%) images are all below 2. For the remaining 120 images, the $\sigma_{noise}$ ranges from 2 to 14.86. By manually analyzing the images, and doing the visual inspection, we find images with $\sigma_{noise}$ over 10 appear to be extremely noisy and distorted, and contains some kinds of perturbations. We find 52 such images, and we use them for our comprehensive analysis of reCAPTCHA's anti-recognition attempts.


We suspect that reCAPTCHA might be using adversarial examples \cite{szegedy2013intriguing, Biggio_2013}, slightly perturbed images maliciously crafted to fool pre-trained image classifiers. Prior work also recommended using adversarial examples to limit the impact DNN image classifier based attacks \cite{Sivakorn_2016}. We experiment to validate our assumption. First, we labeled the potential grids in images with perturbations. We then extract these grids from the images and submit them to the computer vision services for recognition. 

Table \ref{tab:noisy_image_} shows the label sets returned by 4 off-the-shelf image recognition services for an actual reCAPTCHA challenge image. The target object in the challenge is a fire hydrant, and the potential grids are 1, 3, and 4. Table \ref{tab:noisy_grid_classification} shows the result of our experiment. We can see that vision APIs have misclassified a significant number of potential grids. Clarifai API has the highest number of misclassifications, followed by Microsoft Computer Vision API. Google Cloud Vision API did not return any labels for 32 potential grids. Note that we consider a label set to be \textit{acceptable} (A) if at least one of the tags in the set describes the image's content to some extent, even if the name of the target object is not present in it. For example, a potential grid for the target object car may also contain other objects like road, traffic light, and crosswalk. If an image recognition API returns a label set with the tags crosswalk, street, and traffic light, we consider it acceptable even though it does not contain the tag for the main target object (a ``car'' in this case). We consider a label set to be an \textit{exact match} (EM) if one of the returned labels in the set matches the name of the target object. A label set is said to be \textit{misclassified} (MC) if none of the returned labels in the label set has any semantic relation with the grid's actual content. For instance, as shown in Table \ref{tab:noisy_image_}, Google Cloud Vision API misclassified the potential grid number 3, which is an image of a fire hydrant, as an Art (or a Plant). A label set is considered \textit{empty} (E) if the API does not return any label.
In our experiment, we find many grids that are misclassified by image recognition services with very high (over 90\%) probability, which is a strong indication that those grids are adversarial. Generally, non-adversarial perturbation does not mislead a well-trained image classifier; instead, it degrades the target class's confidence score. At the same time,  vision APIs for image recognition return correct or acceptable label sets for some noisy girds (see Table \ref{tab:noisy_grid_classification}). Hence, based on our findings, we hypothesize that reCAPTCHA is using a mixture of adversarial perturbations and random noises in some challenge images.
Note that the actual identification of adversarial perturbations or examples is a non-trivial task, and it is still an open research problem in the AI domain. 

Next, we investigate the impact of adversarial perturbations or random noises on our object detection models. We run our pre-trained object detection models on the same 52 images to determine how many target objects they can correctly identify. Note that, it is not always necessary to detect and localize all the target objects in a challenge image, \textit{i.e.}, it is sufficient to pass a challenge if the \textit{bounding box to grid mapping} algorithm result can capture all the potential grids regardless of the number of objects present in the challenge image. Our base object detection models can recognize less than 60\% of the target objects in the challenge images (see Table \ref{tab:obj_det_perf}). We use the object counts as a metric to assess object detection models' performance to simplify the analysis. 


Next, we apply different data augmentation techniques to study whether retraining the object detection system using the augmented data helps increase the system's robustness against reCAPTCHA's anti-recognition mechanisms. We develop a data augmentation pipeline employing various augmentation methods such as additive Gaussian noise injection, blurring, and changing the brightness and contrast, etc. 
We utilize the \textit{imgaug} \cite{imgaug} library for data augmentation. For adding Gaussian noise, we use the \texttt{AdditiveGaussianNoise} function with $scale$=(0, .2$\ast$255). We use the following methods for blurring: \texttt{GaussianBlur} with $sigma$=(0.5-5.0), \texttt{MedianBlur} with $k$=(5, 17), and \texttt{AverageBlur} with $k$=(5, 17). Figure  \ref{fig:dataAugmentation} in Appendix \ref{appendix:data_aug} shows some examples of data augmentation methods applied to a sample reCAPTCHA challenge image.
%
We randomly select 30\% of training images and apply each data augmentation method from our pipeline to images. Finally, we retrain our object detection models using the augmented training samples. We also collect 500 perturbed images ($\sigma_{noise}$ > 5) from the reCAPTCHA challenges and fine-tune the models further by utilizing \textit{adversarial training}. Note that, even though we call it adversarial training, some training samples may not include adversarial noises. Since reCAPTCHA's source code and data are not open-source, it is challenging to do further verification. 

Table \ref{tab:obj_det_perf} depicts the performance of our object detection models. We can see that the object detection models with augmented data can detect 149 (over 73\%) out of 203 target objects in the perturbed images. That is over 17\% performance improvement with respect to our base models. It is also evident that adversarial training provides significant performance boosts further while using only 500 training samples. We suspect reCAPTCHA might be generating the perturbation from a simple data distribution, which enabled us to achieve such a great increase in performance despite training against only a small number of adversarial samples. We expect that adding more perturbed images from original reCAPTCHA challenges will further enhance the detection performance. Notice that models have misidentified some objects after performing data augmentation and adversarial training. We can reduce the number of misdetections by setting the detection threshold to a higher value (our default is 0.2). However, doing so slightly degrades the overall performance of the object detection models.

We note that advanced object detection systems, such as YOLOv3, are less susceptible to anti-recognition techniques employed by reCAPTCHA in general. For example, our base object detection models (Table \ref{tab:obj_det_perf}) perform much better in identifying objects in the perturbed images than vision APIs for image recognition in Table \ref{tab:noisy_grid_classification}.
We assume that reCAPTCHA mainly targeted image recognition and classification systems because all of the prior attacks against reCAPTCHA in literature are based on them. In summary, our findings imply that an effectively trained object detection based solver can neutralize reCAPTCHA's anti-recognition attempts.

\begin{table}[!t]
\footnotesize
\caption{Results of noisy grid classification returned by image recognition services. EM=No. of exact match label sets. A=No. of acceptable label sets. E=No. of empty label sets. MC=No. of misclassifications label sets.} 
\label{tab:noisy_grid_classification}
\begin{center}
\footnotesize
\begin{tabular}{Lc|cccc}
     \toprule
    \multicolumn{2}{c|}{} & \multicolumn{4}{c}{\textbf{Labels}}\\
    \textbf{Service} &  \textbf{\# of Grids} & \textbf{EM} & \textbf{A} & \textbf{E} & \textbf{MC} \\
    \hline
    Google Cloud Vision & 172 & 31 & 68 & 32 & 41 \\
    \hline
    Microsoft Azure Computer Vision & 172 & 19 & 82 & 9 & 62 \\
    \hline
    Amazon Rekognition & 172 & 70 & 58 & 0 & 44 \\
    \hline
    Clarifai & 172 & 27 & 71 & 0 & 74 \\
    \bottomrule
\end{tabular} 
\end{center}
\end{table}

%
%
%

\paragraph{IP address rate-limit} To study whether reCAPTCHA enforces any IP address rate limit, we set up a 3-day experiment. We select 3 reCAPTCHA-enabled websites and attempt to initiate 1000 reCAPTCHA challenges to a chosen website each day. We limit ourselves to 1000 requests to a site each day to minimize the impact on the test website. Further, there is a delay of 60 seconds between two subsequent requests. We perform this experiment with 3 IP addresses: an institutional IP, a residential IP, and a Tor anonymity network IP. 

We experiment with the academic IP first. On the first day, we can initiate 818 reCAPTCHA challenges, and the remaining 182 attempts have been blocked. On the second day, we are able to initiate the reCAPTCHA challenges 801 times, and the remaining 199 attempts have been blocked. On the third day, none of our attempts has been blocked. The duration of the blocking period usually ranges from 36 to 95 minutes. Note that getting the IP blocked in one website by reCAPTCHA does not generally restrict that same IP from initiating reCAPTCHA challenges on other websites, which is normal behavior. We could initiate reCAPTCHA challenges more than 800 times from a single IP address to a particular website in any of the cases. Next, we repeat this experiment from the same machine but with a residential IP and observe a similar pattern. Finally, we experiment by tunneling the traffic through the Tor network with the exit node in Germany (selected randomly by the Tor client). During this experiment, a significant number of requests have been blocked by reCAPTCHA. Specifically, at least 30\% of our requests are blocked each day. It is worth noting that the exit node's geolocation does not usually make that much of a difference. We confirm this by repeating the experiment separately with a manually specified exit node in three different regions, namely North America (the US), Europe (Netherlands), and Asia (Hong Kong). It implies that reCAPTCHA considers requests originating from an IP within the Tor network to be highly suspicious.

Our findings also indicate that reCAPTCHA's anti-bot technology follows a relaxed per IP address rate-limit approach towards regular IP addresses. While it may be reasonable for a modern web application to allow thousands of requests from a single client machine with a unique IP address, a webpage dedicated for the sole purpose of user registration or login may not want to provide such freedom. Therefore, we recommend letting the website owners set up a custom daily IP address rate-limit by adding such an option in reCAPTCHA deployment settings and enforcing the restriction from the reCAPTCHA backend.

\begin{table}[!t]
\footnotesize
\caption{Impact of anti-recognition on object detection models.} 
\label{tab:obj_det_perf}
\begin{tabular}{C{2.5cm}| C{1cm}| C{1.3cm}| C{2cm}}
\toprule
\textbf{Model} & \textbf{Objects} & \textbf{Objects Detected} & \textbf{Objects Detected with Wrong Label} \\ 
\hline
Base (B) & 203 & 114 & 1 \\
\hline
B+Basic Augmentation (BA) & 203 & 149 & 11 \\
\hline
B+BA+Adversarial Training & 203 & 167 & 13 \\
\bottomrule
\end{tabular} 
\end{table}




\vspace{-2mm}
\section{Economic analysis} \vspace{-2mm}
We use five popular human-based online CAPTCHA solving services to compare their performance with our system. The services are 2Captcha \cite{2captcha}, Anti-Captcha \cite{anticaptcha}, BestCaptchaSolver \cite{bestcaptchasolver}, DeathByCaptcha \cite{deathbycaptcha}, and Imagetyperz \cite{imagetyperz}. We submit 500 reCAPTCHA challenges to each service, totaling 2500 challenges for all the services combined. The average success rate and speed of breaking reCAPTCHA challenges are shown in Table \ref{tab:human_based_solving_perf}. As we can see, our system outperforms both BestCaptchaSolver and Imagetyperz.  
While 2Captcha, Anti-Captcha, and DeathByCaptcha perform a little better than ours, they are significantly slower. Our system is more than 3.5x faster than the fastest human-based CAPTCHA solving service. Further, the adversaries can run our system with virtually no cost by deploying it on their machines. Overall, our system's performance is comparable to that of human-based online CAPTCHA solving services, and scammers could use it as an alternative to human workers to automatically solve reCAPTCHA challenges.

\vspace{-2mm}
\section{Comparing to prior attacks} \vspace{-2mm}
Sivakorn \textit{et al}. leveraged online image annotation APIs to break the earlier implementation of reCAPTCHA (reCAPTCHA 2015) with a success rate of 70.78\% \cite{Sivakorn_2016}. Since then, reCAPTCHA has changed significantly.
In a similar attack, Weng \textit{et al}. evaluated the security of 10 real-world image CAPTCHAs, including reCAPTCHA 2018 \cite{Weng_2019}. They used a CNN-based image classification model to break reCAPTCHA 2018 challenges with a success rate of 79\%.
%
%
Note that reCAPTCHA 2018 used to show relatively simple images when compared to the current reCAPTCHA challenges. Further, Weng \textit{et al}. encountered only 10 image categories in the reCAPTCHA 2018 challenges, where we come across 18 object categories in the latest version of reCAPTCHA. Moreover, the anti-recognition mechanism employed by the current reCAPTCHA was not available in reCAPTCHA 2018 as well. 

In summary, we propose a new approach to breaking the most advanced version of reCAPTCHA using object detection models. 
Our method significantly outperforms prior approaches as well as off-the-shelf object detection APIs.
%
%
We believe the stark difference in the performance between our solver and off-the-shelf object detection APIs is because we train our solver to handle reCAPTCHA object categories exclusively. In contrast, object detection APIs are developed for general-purpose object detection tasks.  
Further, as discussed before, we assume that these services are still in their early development stages. 

\begin{table}[!t]
\centering
\small
    \caption{Performance of human-based CAPTCHA solving services. 
    } 
    \label{tab:human_based_solving_perf}
    \begin{tabular}{c|c|c}
    \toprule
     \textbf{Service} & \textbf{Success Rate (\%)} & \textbf{Speed (s)} \\
     \hline
     2Captcha & 98.2 & 73.11 \\  
     \hline
     Anti-Captcha & 92.4 & 83.99 \\
     \hline
     BestCaptchaSolver & 67.2 & 93.42 \\
     \hline
     DeathByCaptcha & 96.2 & 78.33 \\
     \hline
     Imagetyperz & 73 & 131.4 \\
     \hline
     \hline
     Our system & 83.25 & 19.93 \\
     \bottomrule
    \end{tabular} 
\end{table}

\vspace{-2mm}
\section{Discussion} \vspace{-2mm}

\subsection{Ethics}
We did not affect the security or the availability of the tested websites during our data collection for preliminary analysis or performing a live attack on reCAPTCHA as we limit our access within the two \texttt{iframe} elements related to the challenge.
We also disclosed our findings to Google when we developed our system's initial implementation in August 2019. Unfortunately, we have not noticed any discernible changes to reCAPTCHA by Google that can prevent our attack. Our system can still break the reCAPTCHA challenges with a high success rate as of March 2020.  

We have not published the source code of our tool due to concerns over potential abuse by scammers and fraudsters alike. However, we encourage researchers to contact us if they want to use our tool for research purposes only.

\subsection{Limitation}
We design our attack to break reCAPTCHA challenges specifically. While reCAPTCHA is the most widely deployed CAPTCHA service on the Web, there are other popular image CAPTCHA schemes. It will be interesting to see if we could extend our object detection based solver module to attack a whole family of similar image CAPTCHA designs. We plan to conduct a study on the generalization of our attack as a future extension.




\subsection{Countermeasures} 
While it may not be possible to prevent our attack completely, we provide several countermeasures to limit it.


\paragraph{Content heterogeneity}
Our experiment shows that content homogeneity has contributed to lowering the accuracy of image recognition and classification services. However, it has minimal to no impact on our object detection based solver. As such, reCAPTCHA's current approach to resisting automated attacks does not seem to be working. We recommend using images from diverse and heterogeneous sources, which will provide the CAPTCHA designers more flexibility if they need to expand the total number of object categories.  

\paragraph{Incorporate natural language understanding to image CAPTCHA test} The natural language understanding is considered as one of the three biggest open problems in natural language processing \cite{talk2018}. This weakness could be exploited to strengthen the security of image CAPTCHA. We suggest utilizing the natural language understanding in forming the challenge instruction so that the direction needed to solve a challenge must be inferred through natural language reasoning. The current design of reCAPTCHA makes this information readily available to the attacker. 

\paragraph{Use spatial properties of the object}
The main design flaw of reCAPTCHA is that an advanced object detection system can solve its underlying AI problem for telling humans and bots apart. The problem could be hardened for the machine by exploiting the object's spatial attributes, such as shape, size, orientation, tilt direction, etc. However, it may require extensive research to determine whether designing such a CAPTCHA scheme is feasible in practice.

\vspace{-1mm}
\section{Related work} \vspace{-1mm}
CAPTCHA is an active research area, and there exists an extensive body of studies in this area. Due to space limitations, we only discuss the works that are mostly related to ours. Further, we mainly focus on CAPTCHA attack related research.  

\paragraph{Image CAPTCHAs} 
Golle \textit{et al}. \cite{Golle_2008} used support vector machine classifiers to break Asirra CAPTCHA \cite{elson2007asirra}. Zhu \textit{et al}. analyzed the security of various earlier image CAPTCHAs and proposed attacks to break them \cite{Zhu_2010}. 
Sivakorn \textit{et al}. used deep learning techniques to break reCAPTCHA 2015 \cite{Sivakorn_2016}. Later Weng \textit{et al}. analyzed the security of several real-world image CAPTCHAs, including reCAPTCHA 2018, and developed deep learning-based attacks that succeeded in breaking all the CAPTCHAs tested in their work \cite{Weng_2019}.  Osadchy \textit{et al}. proposed a new CAPTCHA scheme called DeepCAPTCHA that exploits adversarial examples in CAPTCHA image generation to deceive DNN image classifiers \cite{Osadchy_DeepCAPTCHA2017}. Shi \textit{et al}. proposed a framework for generating text and image adversarial CAPTCHAs \cite{shi2019adversarial}. 

\paragraph{Text CAPTCHAs} 
Most text CAPTCHA schemes have been broken \cite{Mori:2003:ROA:1965841.1965858, Moy:2004:DET:1896300.1896305, Chellapilla:2004:UML:2976040.2976074, Yan2007BreakingVC, Yan_2008, Gao2016ASG}.
Chellapilla \textit{et al}. proposed using machine learning algorithms to break earlier text CAPTCHA designs \cite{Chellapilla:2004:UML:2976040.2976074}. Yan \textit{et al}. used simple pattern recognition algorithms to break most of the text CAPTCHAs provided at Captchaservice.org with a near-perfect success rate \cite{Yan2007BreakingVC}.
El Ahmad \textit{et al}. proposed a novel attack against reCAPTCHA v1 2010 \cite{elahmad2011robustness}.
In 2011, Bursztein \textit{et al}. evaluated the security of 15 CAPTCHA schemes from popular web sites and concluded that 13 of them were vulnerable to automated attacks \cite{Bursztein_CCS11}.
In 2014, Bursztein \textit{et al}. used a machine learning-based generic attack to break many popular real-world text CAPTCHA schemes, including reCAPTCHA 2011 and reCAPTCHA 2013 \cite{Bursztein_2014}. In 2016, Gao \textit{et al}. were able to break many text CAPTCHAs using a low-cost attack that uses Log-Gabor filters \cite{Gao2016ASG}. In 2018, Ye \textit{et al}. proposed a Generative Adversarial Networks (GANs) based approach to break 33 text CAPTCHA schemes \cite{Ye_2018}.

\paragraph{Audio CAPTCHAs}
Audio CAPTCHAs designed as alternative CAPTCHA schemes for visually impaired users have been subjected to automated attacks over the years \cite{Bursztein_2009, Tam_NIPS2008, Sano2015JIP, Bock:2017:ULD:3154768.3154775, Solanki2017_audio}. 
Tam \textit{et al}. analyzed the security of audio CAPTCHAs from popular websites by using machine learning techniques and were able to break many of them, including an earlier version reCAPTCHA \cite{Tam_NIPS2008}.
In 2017, Bock \textit{et al}. proposed an automated system called unCaptcha that could break reCAPTCHA's audio challenges with an accuracy above 85\% \cite{Bock:2017:ULD:3154768.3154775}. In 2017, Solanki \textit{et al}. proposed a low-cost attack against several popular audio CAPTCHAs using off-the-shelf speech recognition services \cite{Solanki2017_audio}.

\vspace{-3mm}
\section{Conclusion} 
CAPTCHAs have become a standard security mechanism for protecting websites from abusive bots. 
In this work, we showed that one of the Internet's most widely used image CAPTCHA schemes, reCAPTCHA v2, could be broken by an automated solver based on object detection method with a high success rate. Our extensive analysis showed that despite several major security updates to counter automated attacks, which could invalidate prior image recognition and classification based solvers, reCAPTCHA remains vulnerable to advanced object detection systems.
%
%
Given the capabilities of the current object detection systems, we think that reCAPTCHA is essentially broken because its reverse Turing tests to distinguish humans from bots are easily solvable by an object detection based automated solver. 

\section*{Acknowledgment}
This work is partially supported in part by US NSF under grants OIA-1946231. The work in China is supported by Cooperation and Exchange Program of International Science and Technology of Shaanxi Province (2019KW-010). We also want to thank our shepherd Kevin Borgolte and the anonymous
reviewers for valuable comments.

\bibliographystyle{plain}
\bibliography{main}



\begin{appendices}
\section{Captcha Types} \label{appendix:captchaTypes}
\begin{figure}[!htp]
\centering
     \includegraphics[width=0.60\linewidth, height=6cm]{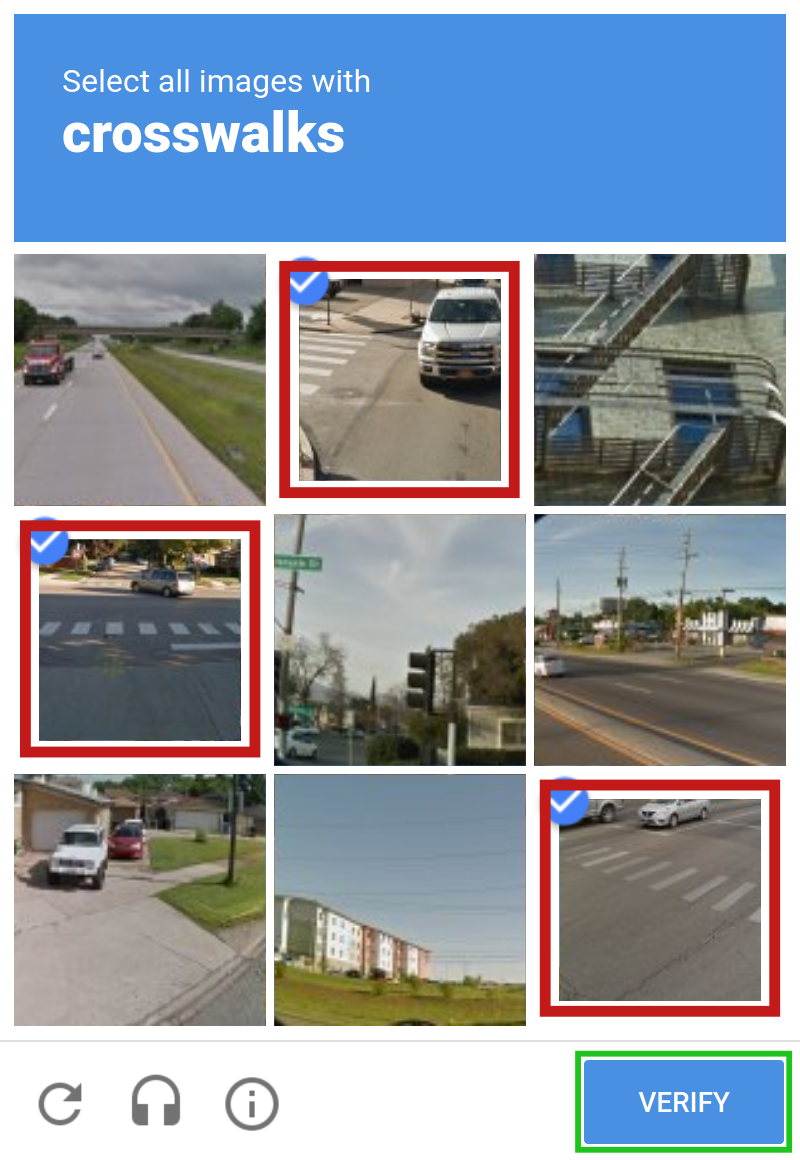}
     \caption{Selection-based image CAPTCHA.} 
 \label{fig:sel_captcha}
\end{figure}

\begin{figure*}[!tp] 
\centering
 \includegraphics[width= 0.83\linewidth,height=6cm]{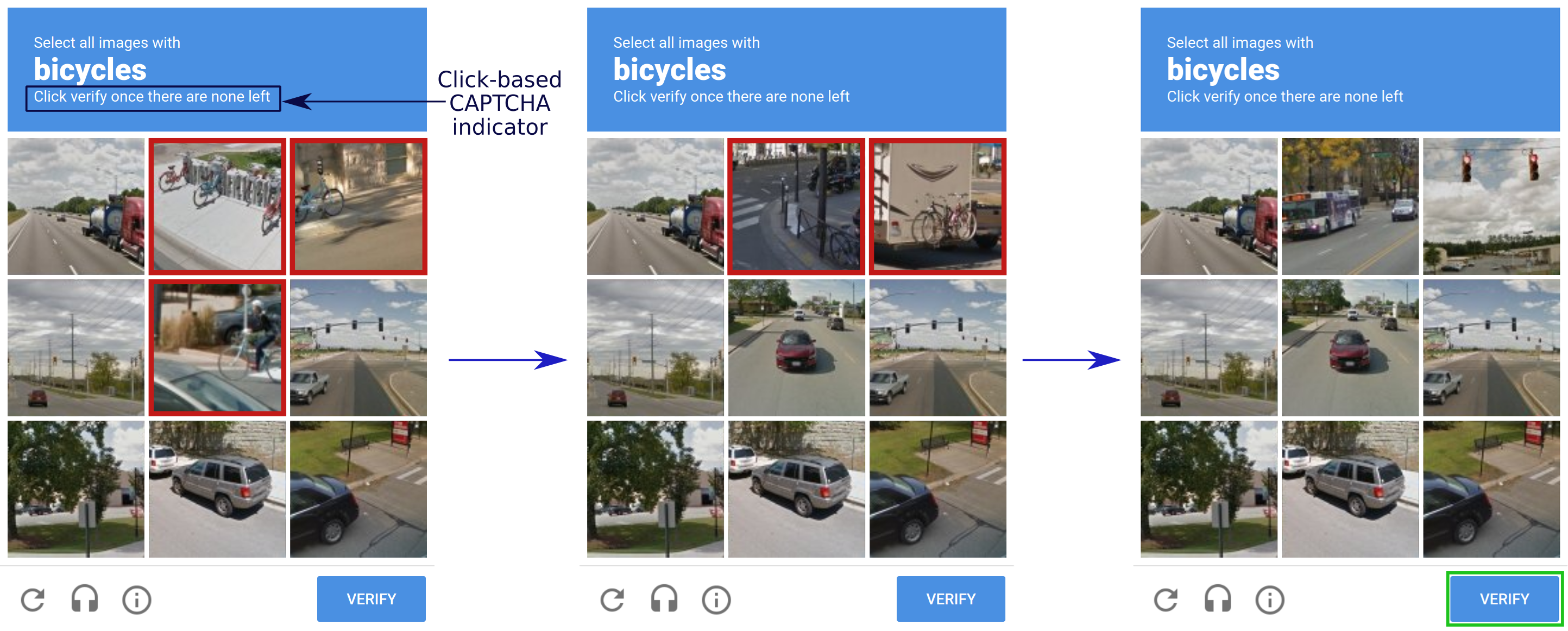}
 \caption{Click-based image CAPTCHA.}
 \label{fig:click_captcha} 
\end{figure*}

Figure \ref{fig:sel_captcha} shows an example of a selection-based CAPTCHA challenge, and Figure \ref{fig:click_captcha} shows an example of a click-based CAPTCHA challenge. 

\begin{figure*}[!htp]
\centering
    \begin{subfigure}{.42\textwidth}
        \centering
        \includegraphics[width=.9\linewidth, height=6cm]{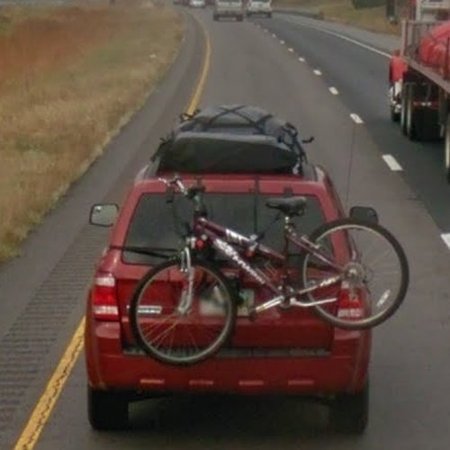}
        \caption{Original image}
    \end{subfigure}
    \begin{subfigure}{.42\textwidth}
        \centering
        \includegraphics[width=.9\linewidth, height=6cm]{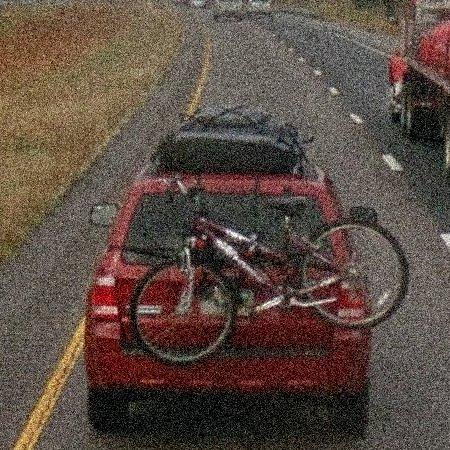}
        \caption{AdditiveGaussianNoise ($scale$=.1$\ast$255)}
    \end{subfigure} \vspace{2mm}

    \begin{subfigure}{.28\textwidth}
        \centering
        \includegraphics[width=.85\linewidth]{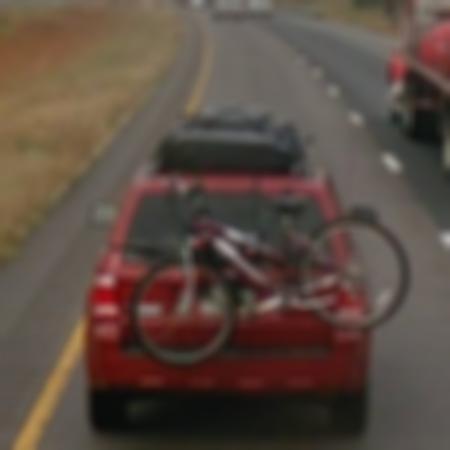}
        \caption{GaussianBlur ($sigma$=5.0)}
    \end{subfigure}
    \begin{subfigure}{.28\textwidth}
        \centering
        \includegraphics[width=.85\linewidth]{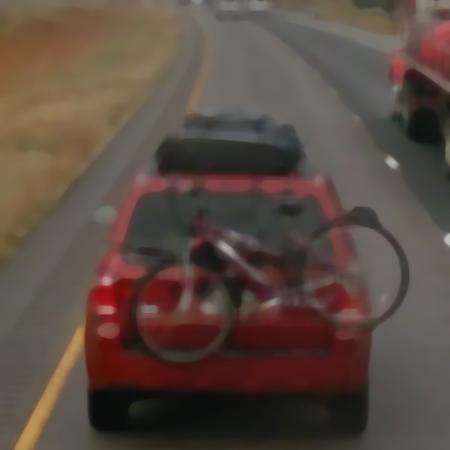}
        \caption{MedianBlur ($k$=13)}
    \end{subfigure}
    \begin{subfigure}{.28\textwidth}
        \centering
        \includegraphics[width=.85\linewidth]{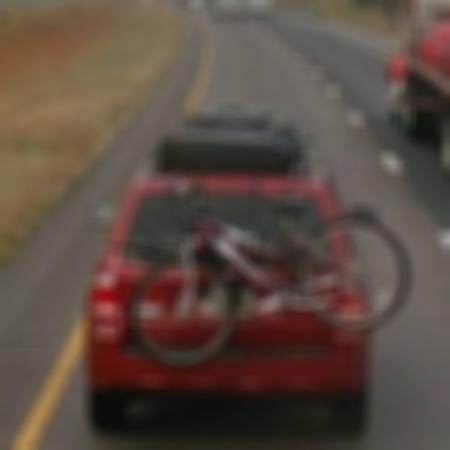}
        \caption{AverageBlur ($k$=15)}
    \end{subfigure}
\caption{Examples of data augmentation methods.}
\label{fig:dataAugmentation}
\end{figure*}

\section{Data Augmentation} \label{appendix:data_aug}
Figure \ref{fig:dataAugmentation} depicts some examples of data augmentation methods applied to a sample reCAPTCHA challenge image.
\end{appendices}

\end{document}